\documentstyle[12pt,graphicx,subfigure]{article}
\def\Journal#1#2#3#4{{#1} {\bf #2}, #3 (#4)}

\def\NPB{{\em Nucl. Phys.} B}
\def\PLB{{\em Phys. Lett.}  B}
\def\PRL{\em Phys. Rev. Lett.}
\def\PRD{{\em Phys. Rev.} D}

\def\beq{\begin{equation}}
\def\eeq{\end{equation}}
\def\lsim{\ ^<\llap{$_\sim$}\ }
\def\gsim{\ ^>\llap{$_\sim$}\ }
\def\r2{\sqrt 2}
\def\beq{\begin{equation}}
\def\eeq{\end{equation}}
\def\beqn{\begin{eqnarray}}
\def\eeqn{\end{eqnarray}}

\def\sinW2{\sin^2\theta_W}

\def\mz2{M_{z}^2}
\def\c2b{\cos 2\beta}

\def\mz{M_z}

\def\Fq2{F_{2}(q^2)}

\def\sec2w{sec^2\theta_W}

\def\gmin2{(g-2)_\mu}

\catcode`\@=11 

\def\lsim{\mathrel{\mathpalette\@versim<}}
\def\gsim{\mathrel{\mathpalette\@versim>}}
\def\@versim#1#2{\vcenter{\offinterlineskip
    \ialign{$\m@th#1\hfil##\hfil$\crcr#2\crcr\sim\crcr } }}

\def\PRL{Phys. Rev. Lett.}

\begin{document}

\begin{titlepage}

\begin{center}
{\large {\bf Effective Lagrangian for $\chi^+_i\chi_j^0H^-$ Interaction 
  in the MSSM and Charged Higgs Decays}}\\
\vskip 0.5 true cm
\vspace{2cm}
\renewcommand{\thefootnote}
{\fnsymbol{footnote}}
 Tarek Ibrahim$^{a,b}$, Pran Nath$^{b}$ and Anastasios Psinas $^{b}$  
\vskip 0.5 true cm
\end{center}

\noindent
{a. Department of  Physics, Faculty of Science,
University of Alexandria,}\\
{ Alexandria, Egypt\footnote{: Permanent address of T.I.}}\\ 
{b. Department of Physics, Northeastern University,
Boston, MA 02115-5000, USA } \\
\vskip 1.0 true cm
\centerline{\bf Abstract}
\medskip
We extend previous analyses of the supersymmetric loop correction to the
charged Higgs couplings to include the coupling $H^{\pm}\chi^{\mp}\chi^0$. 
The analysis completes the previous analyses where similar corrections 
were computed for $H^+\bar t b$ ($H^-t\bar b$), and for
$H^+\tau^- \bar\nu_{\tau}$ ($H^- \tau^+ \nu_{\tau}$)
 couplings within the minimal supersymmetric
standard model.  The effective one loop Lagrangian is then applied to
the computation of the charged Higgs decays. 
 The sizes of the supersymmetric loop correction on branching ratios of 
the charged Higgs $H^+(H^-)$ into the decay modes 
 $t\bar b$ ($\bar t b$), $\bar\tau\nu_{\tau}$ 
($\tau \bar \nu_{\tau}$),  and 
$\chi_i^+\chi_j^0(\chi_i^-\chi_j^0)$ (i=1,2; j=1-4) are investigated and the 
supersymmetric loop correction is found to be significant, i.e., in the range
20-30\% in significant regions of the parameter space.
The loop correction to the decay mode $\chi_1^{\pm}\chi_2^0$ is examined
in specific detail as this decay mode leads to a trileptonic signal.
 The effects of CP phases on the branching ratio are also 
investigated. A brief discussion of the implications of the
analysis for collders is given.
 \end{titlepage}
\section{Introduction}
The Higgs couplings to matter and to gauge fields are of great
current interest as they enter in a variety of phenomena which
are testable in low energy processes\cite{carena2002}. 
Specifically it has been
known for some time that the loop correction to the b quark 
mass generates a contribution  which becomes large for large
$\tan\beta$ underlining the importance of the loop correction in
phenomena involving the Higgs boson couplings\cite{susybtmass}. Recently 
analyses of the supersymmetric one loop corrections to the Higgs boson
 couplings  were given and its implications for 
the decay of the Higgs into $H^+\rightarrow t\bar b$
($H^-\rightarrow \bar t b$) and $H^+\rightarrow \bar \tau\nu_{\tau}$
($H^-\rightarrow \tau\bar\nu_{\tau}$)
were 
analysed\cite{Carena:1999py,Christova:2002sw,Ibrahim:2003ca,Ibrahim:2003jm,Ibrahim:2003tq} . 
These decays
are of great importance as they differ strongly from the predictions 
in the Higgs sector of the Standard Model and thus provide possible signals for
the observation of supersymmetry at colliders. In the analysis
given in Refs.\cite{Christova:2002sw,Ibrahim:2003ca,Ibrahim:2003jm,Ibrahim:2003tq}
 the decay of the Higgs into chargino and neutralinos
was, however, not considered. In this paper we extend the analysis
to include the loop correction to the 
$H^{\pm}\chi^{\mp}\chi^0$
couplings. We also take into account the effects of the CP phases.

The analysis is carried out in the framework of the minimal supersymmetric
standard model (MSSM). For the numerical part of the analysis we work
within the framework of extended supergravity unified models. 
Thus the mimimal supergravity unified model (mSUGRA)\cite{msugra} 
is parametrized 
by the universal scalar mass $m_0$, the universal gaugino mass
$m_{\frac{1}{2}}$,  the universal trilinear coupling $A_0$, 
the ratio of the Higgs vacuum expectation values (VEVs), i.e., 
 $\tan\beta =<H_2>/<H_1>$ where $H_2$ gives mass to the up quark and
 $H_1$ gives VEV to the down quark  and the lepton, and sign($\mu)$) where
 $\mu$ is the Higgs mixing parameter which appears in the
 superpotential in the form $\mu H_1H_2$.  
 mSUGRA is based on the assumption of a flat Kahler potential and thus
 can be extended by inclusion of more general Kahler potentials.
 This allows one to introduce nonuniversalities in the soft
 parameters. Thus for more general analyses,we will assume 
 nonuniversalities in the Higgs sector, and also allow for CP phases.
 The inclusion of phases of course involves attention to the
 severe experimental constraints that exist on the electric dipole 
 moment (edm) of the electron\cite{eedm}, of the neutron\cite{nedm}  
 and of $^{199}Hg$ atom\cite{atomic}. 
 However, as is now well known there are a variety of techniques 
 available that allow one to suppress the large edms  and bring them
 in conformity with the current 
 experiment\cite{na,incancel,olive,chang}.
 CP phases  affect loop corrections to the Higgs mass\cite{cphiggsmass},
 dark matter\cite{cpdark} and  a  number of other phenomena
(for a review see  Ref.\cite{Ibrahim:2002ry}).
The outline of the rest of the paper is as follows: In Sec.2 we 
compute the loop correction to the $H^{\pm}\chi^{\mp}\chi^0$ couplings
arising from supersymmtric particle exchanges and the effects
of these corrections on the charged Higgs decay. In Sec.3 we give
a numerical analysis of the sizes of radiative corrections.
 It is found that the loop
correction can be  as  large as 25-30\% in certain parts of 
the parameters space. Implications of these results at colliders
are briefly discussed in Sec.4 and conclusions are given in Sec.5.

\section{Loop Corrections to Charged Higgs Couplings}
The microscopic Lagrangian for $H^{\pm}\chi^{\mp}\chi^0$ interaction is
\beqn
{\cal {L}}=\xi_{ji}H_2^{1*}\bar \chi_j^0P_L \chi_i^+ 
+ \xi_{ji}'H_1^2\bar \chi_j^0P_R \chi_i^+ +H.c.
\label{hchichicouplings}
\eeqn
where $H_1^2$ and $H_2^1$ are the charged states of the two Higgs 
iso-doublets in the minimal supersymmetric standard model (MSSM),
.i.e, 
\beqn
(H_1)= \left(\matrix{H_1^1\cr
 H_{1}^2}\right),~~
(H_2)= \left(\matrix{H_{2}^1\cr
             H_2^2}\right)
\label{a}
\eeqn
and $\xi_{ji}$ and $\xi_{ji}'$ are given by 
\beqn
\xi_{ji}=-gX_{4j}V_{i1}^* 
-\frac{g}{\sqrt 2} X_{2j} V_{i2}^* 
-\frac{g}{\sqrt 2} \tan\theta_WX_{1j}V_{i2}^*
\eeqn
and 
\beqn
\xi_{ji}'=-gX_{3j}^*U_{i1} 
+\frac{g}{\sqrt 2} X_{2j}^* U_{i2} 
+\frac{g}{\sqrt 2} \tan\theta_WX_{1j}^*U_{i2}
\eeqn
where $X$, $U$ and $V$ diagonalize  the neutralino and chargino mass matrices 
so that
\beqn
X^T M_{\chi^0}X= diag (m_{\chi_1^0},
m_{\chi_2^0},m_{\chi_3^0},m_{\chi_4^0})\nonumber\\
U^* M_{\chi^+}V^{-1}= diag (m_{\chi_1^+}, m_{\chi_2^+})
\eeqn
where $m_{\chi_i^0}$ (i=1,2,3,4) are the eigen values of the neutralino
mass matrix $M_{\chi^0}$ and $m_{\chi_1^+}, m_{\chi_2^+}$ are the
eigen values of the chargino mass matrix $M_{\chi^+}$.

The loop corrections produce shifts in the couplings of 
Eq.~(\ref{hchichicouplings}) and the effective  Lagrangian with loop corrected
couplings is given by

\beqn
{\cal {L}}_{eff}=(\xi_{ji} + \delta\xi_{ji} ) H_2^{1*}\bar \chi_j^0P_L \chi_i^+
+ \Delta\xi_{ji} H_1^2\bar\chi_j^0 P_L \chi_i^+ \nonumber\\
+ (\xi_{ji}' +\delta \xi_{ji}') H_1^2\bar\chi_j^0P_R \chi_i^+ +
\Delta \xi_{ji}' H_2^{1*}\bar \chi_j^0P_R \chi_i^+ + H.c.
\label{hchichicouplings1}
\eeqn
In this work we calculate the loop correction to the 
$\chi^{\pm}\chi^0H^{\mp}$ using the zero external momentum 
approximation.
\begin{figure}
\hspace*{-0.6in}
\centering
\includegraphics[width=8cm,height=4cm]{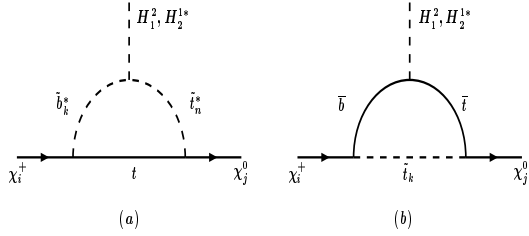}
\caption{The stop and sbottom exchange contributions to the
$H^-\chi^+\chi^0$ vertex.}                                                                              		  	                              
\label{figab}
\end{figure}
\subsection{Loop analysis of $\Delta\xi_{ij}$} 
The corrections to $\Delta\xi_{ij}$ in the zero extrenal momentum 
approximation arise from the loop diagrams 
Figs.(1)-(4) so that 
\beqn
\Delta\xi_{ji}= \Delta\xi_{ji}^{(1a)} + \Delta\xi_{ji}^{(1b)} +
\Delta\xi_{ji}^{(2a)} +\Delta\xi_{ji}^{(2b)}+ 
\Delta\xi_{ji}^{(3a)}+   \Delta\xi_{ji}^{(3b)}
+\Delta\xi_{ji}^{(4)}
\label{Deltaxi}
\eeqn
 We  note that the contribution from diagrams which have 
 $W-Z-\chi_i^0$ and $W-Z-\chi_i^+$ exchanges in the loop 
 vanish due to the absence of $H^+W^-Z$
vertex at tree level. This is a general feature of models with two
doublets of Higgs\cite{gm}.
Also the loops with $H^+W^-H^0_k$ and
$H^+ZH^-$ vertices do not contribute in the zero external
momentum approximation since these vertices are proportional to the external 
momentum. Since we wish to apply the effective couplings to 
the decay of the charged Higgs into charginos and neutralinos, 
the mass of the charged Higgs must be relatively large.
 Thus we have ignored the other diagrams which have  $H^{\pm}$ running in the loops due
to the large mass suppression.
 We give now the computation for each of Figs.(1)-(4).

\noindent
{Loop Fig.(1a)}: For the evluation of $\Delta\xi_{ij}$  for Fig. (1a)
we need $\tilde{b}t\chi^{0}$,  $\tilde{t}t\chi^{0}$ and
$\tilde{b}\tilde{t}H$ interactions. These are given by
\begin{equation}
{\cal {L}}_{\tilde{b}t\chi^{+}}=-g\bar{t}[(U_{l1}D_{b1n}-K_{b}U_{l2}D_{b2n})
P_{R}-K_{t}V^{*}_{l2}D_{b1n}P_{L}]\tilde{\chi}^+_l\tilde{b_n}+H.c.
\end{equation}
\begin{equation}
{\cal {L}}_{\tilde{t}t\chi^0}=-\sqrt{2}\bar{t}[(\alpha_{tl}D_{t1n}-\gamma_{tl}D_{t2n})P_L+
(\beta_{tl}D_{t1n}+\alpha^*_{tl}D_{t2n})P_R]\tilde{\chi}^0_l\tilde{t_n}+H.c
\end{equation}
\begin{equation}
{\cal {L}}_{H\tilde{t}\tilde{b}}=H^1_2\tilde{b_k}\tilde{t^*_n}\eta_{kn}+H^2_1\tilde{b^*_k}\tilde{t_n}\eta'_{kn}+H.c
\end{equation}
\beqn
\alpha_{tk} =\frac{g m_tX_{4k}}{2m_W\sin\beta}\nonumber\\
\beta_{tk}=eQ_tX_{1k}^{'*} +\frac{g}{\cos\theta_W} X_{2k}^{'*}
(T_{3t}-Q_t\sin^2\theta_W)\nonumber\\
\gamma_{tk}=eQ_t X_{1k}'-\frac{gQ_t\sin^2\theta_W}{\cos\theta_W}
X_{2k}'
\eeqn
where $X'$'s are given by 
\beqn
X'_{1k}=X_{1k}\cos\theta_W +X_{2k}\sin\theta_W\nonumber\\
X'_{2k}=-X_{1k}\sin\theta_W +X_{2k}\cos\theta_W
\eeqn
and where 
\beqn
K_{t(b)}=\frac{m_{t(b)}}{\sqrt 2 m_W \sin\beta (\cos\beta)}
\eeqn
\begin{figure}
\hspace*{-0.6in}
\centering
\includegraphics[width=8cm,height=4cm]{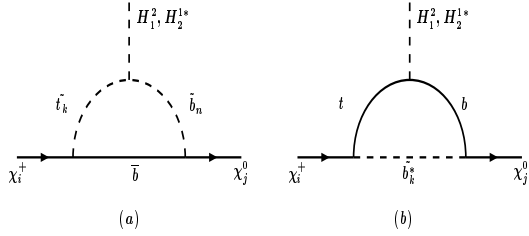}
\caption{Another set of  diagrams exhibiting  stop and sbottom exchange 
contributions to the $H^-\chi^+\chi^0$ vertex. }                                                                              		  	                              
\label{figcd}
\end{figure}
Finally, $\eta_{ij}$ is defined by 
\beqn\label{f}
\eta_{ij}= \frac{gm_t}{\sqrt 2 m_W \sin\beta} m_0A_tD_{b1i} D_{t2j}^*
+\frac{gm_b}{\sqrt 2 m_W \cos\beta}\mu D_{b2i} D_{t1j}^*\nonumber\\
+\frac{gm_bm_t}{\sqrt 2 m_W \sin\beta} D_{b2i} D_{t2j}^*
+\frac{gm_t^2}{\sqrt 2 m_W \sin\beta} D_{b1i} D_{t1j}^*
-\frac{g}{\sqrt 2} m_W \sin\beta  D_{b1i} D_{t1j}^*
\eeqn
and $\eta_{ij}'$ is defined by 
\beqn\label{e}
\eta_{ji}'= \frac{gm_b}{\sqrt 2 m_W \cos\beta} m_0A_bD_{b2j}^* D_{t1i}
+\frac{gm_t}{\sqrt 2 m_W \sin\beta} \mu D_{b1j}^* D_{t2i}\nonumber\\
+\frac{gm_bm_t}{\sqrt 2 m_W \cos\beta}  D_{b2j}^* D_{t2i}
+\frac{gm_b^2}{\sqrt 2 m_W \cos\beta} D_{b1j}^* D_{t1i}
-\frac{g}{\sqrt 2} m_W \cos\beta  D_{b1j}^* D_{t1i}
\eeqn
where
$D_{bij}$ is the matrix that diagonalizes the 
b squark $mass^2$ matrix so that
\beqn
\tilde b_L=\sum_{i=1}^{2} D_{b1i} \tilde b_i,~~~~~ 
\tilde b_R=\sum_{i=1}^{2} D_{b2i} \tilde b_i
\eeqn
where $\tilde b_i$ are the b squark mass eigen states.
In a similar fashion  $D_{tij}$ diagonalizes the t squark 
$mass^2$ matrix so that 
\beqn
\tilde t_L=\sum_{i=1}^{2} D_{t1i} \tilde t_i,~~~~~ 
\tilde t_R=\sum_{i=1}^{2} D_{t2i} \tilde t_i
\eeqn
where $\tilde t_i$ are the t squark mass eigen states. 
Using the above one finds for Fig. (1a) the result
\begin{equation}
\Delta\xi^{(1a)}_{ji}=-\sum_{k=1}^2\sum_{n=1}^2\sqrt{2}gK_tV^*_{i2}D_{b1k}\eta'_{kn}
(\beta^*_{tj}D^*_{t1n}+\alpha_{tj}D^*_{t2n})(\frac{m_t}{16\pi^2})
f(m^2_t,{m}^2_{\tilde b_k},{m}^2_{\tilde t_n})
\end {equation}
where the form factor $f(m^2,m_i^2,m_j^2)$ is defined for $i\neq j$ so that 
\beqn\label{h}
f(m^2,m_i^2,m_j^2)
= \frac {1}{(m^2-m_i^2) (m^2-m_j^2)(m_j^2-m_i^2)}\nonumber\\
(m_j^2 m^2 ln\frac{m_j^2}{m^2} 
 +m^2 m_i^2ln\frac{m^2}{m_i^2} +m_i^2 m_j^2 ln\frac{m_i^2}{m_j^2})  
 \eeqn
and for the case $i= j$ it is given by  
 \beqn\label{i}
 f(m^2,m_i^2,m_i^2) =\frac {1}{(m_i^2-m^2)^2} (m^2 ln\frac{m_i^2}{m^2} 
 + (m^2-m_i^2))
\eeqn 

\noindent
{ Loop Fig.(1b)}:  For  this loop analysis we  need the $Htb$ interaction
\begin{equation}
 {\cal {L}}_{Htb}=\frac{gm_t}{\sqrt{2}m_W\sin{\beta}}\overline{t}P_{L}bH^1_2+\frac{gm_b
}{\sqrt{2}m_{W}\cos{\beta}}\overline{t}P_{R}b{H^2}^{*}_1+H.c
\end{equation}
Using the above interaction along with $L_{\tilde{t}b\chi^+}$ where 
\begin{equation}
{\cal {L}}_{\tilde{t}b\chi^+}=-g\overline{b}[(V_{l1}D_{t1n}-K_{t}V_{l2}
D_{t2n})P_R-K_{b}U^*_{l2}D_{t1n}P_{L}]\tilde{\chi^c_l}\tilde{t_n}+H.c
\end{equation}
one finds the loop correction from Fig.(1b) so that 
\begin{equation}
\Delta\xi^{(1b)}_{ji}=\sum_{k=1}^2\frac{g^2m_b}{m_w\cos{\beta}}[\alpha_{tj}D_{t1k}-\gamma_{tj}D_{t2k}]
[V^*_{i1}D^*_{t1k}-K_{t}V^*_{i2}D^*_{t2k}]\frac{m_{t}m_{b}}{16\pi^2}f(m^2_t,m^2_b,m^2_{\tilde{t_k}})
\end{equation}

\noindent
{ Loop Fig.(2a)}: The analysis for this graph requires in addition the 
$\bar b b\chi^0$ interaction, i.e., 
\begin{equation}
{\cal {L}}_{\tilde{b}b\chi^0}=-\sqrt{2}\overline{b}[(\alpha_{bl}D_{b1n}-\gamma_{bl}D_{b2n})P_L+
(\beta_{bl}D_{b1n}+\alpha^*_{bl}D_{b2n})P_{R}]\tilde{\chi^0_l}\tilde{b_n}+H.c.
\end{equation}
where 
\beqn
\alpha_{bk}=\frac{gm_bX_{3k}}{2m_W \cos\beta}\nonumber\\
\beta_{bk}= eQ_b X^{'*}_{1k} +\frac{g}{\cos\theta_W}X^{'*}_{2k}
(T_{3b}-Q_b \sin^2\theta_W)\nonumber\\
\gamma_{b k} =e Q_b X'_{1k} -\frac{gQ_b\sin^2\theta_W}{\cos\theta_W}
X_{2k}'
\eeqn
The analysis then gives 
\begin{equation}
\Delta\xi^{(2a)}_{ji}=\sum_{k=1}^2\sum_{n=1}^2\sqrt{2}g(V^*_{i1}D^*_{t1k}-K_{t}V^*_{i2}D^*_{t2k})\eta'_{nk}
(\alpha_{bj}D_{b1n}-\gamma_{bj}D_{b2n})\frac{m_b}{16\pi^2} 
f(m^2_b,m^2_{\tilde{t_k}},m^2_{\tilde{b_n}})
\end{equation}

\noindent
{ Loop Fig.(2b)}: Using the interactions of $\overline{b}t\chi^+$,   
$\tilde{b}b\chi^0$, and $Hbt$, one finds 
\begin{equation}
\Delta\xi^{(2b)}_{ji}=-\sum_{k=1}^2g^2K_{t}V^*_{i2}D_{b1k}\frac{m_b}{m_{W}\cos{\beta}}
(\beta^*_{bj}D^*_{b1k}+\alpha_{bj}D^*_{b2k})\frac{m_{t}m_{b}} 
{16\pi^2}f(m^2_{b},m^2_{t},m^2_{\tilde{b_k}})
\end{equation}
\noindent
{ Loop Fig.(3a)}:
For  the loop diagram of Fig.(3a) we need $\chi^{\pm}_{l}\chi^0_{m}W^{\mp}$  and   
$H^{\pm}\chi^0_{l}\chi^{\mp}_{m}$  interactions. 
The  $H^{\pm}\chi^0\chi^{\mp}$ is given by $Eq.(1)$ while the $\chi^+\chi^0 W$ 
interaction is given by 
\begin{equation}
{\cal {L}}_{\chi^{\pm}\chi^0W^{\mp}}=gW^-_{\mu}\overline{\chi^0_i}\gamma^{\mu}
[L_{ij}P_L+R_{ij}P_{R}]\chi^+_j+gW^+_{\mu}\overline{\chi^+_j}
\gamma^{\mu}[L^*_{ij}P_L+R^*_{ij}P_R]\chi^0_i. 
\end{equation}
where
\begin{equation}
L_{ij}=-\frac{1}{\sqrt{2}}X^*_{4i}V^*_{j2}+X^*_{2i}V^*_{j1}
\end{equation}
and 
\begin{equation}
R_{ij}=\frac{1}{\sqrt{2}}X_{3i}U_{j2}+X_{2i}U_{j1}
\end{equation}
 Our metric is such that
 $g_{\mu\nu}\gamma^{\mu}\gamma^{\nu}=4$, and using it one finds   
\begin{equation}
\Delta\xi^{(3a)}_{ji}=-\sum_{m=1}^2\sum_{l=1}^44g^2R^*_{jm}L_{li}
{\xi'_{lm}}\frac{m_{\chi^0_l}m_{\chi_m^-}}{16\pi^2}
f(m^2_{\chi^0_l},m^2_{\chi_m^-},m^2_W)
\end{equation}

\noindent
{ Loop Fig. (3b)}: 
Here we need  the interactions of  $Z\chi^+\chi^-$
and $Z\chi^0\chi^0$
which are given by
\begin{equation}
{\cal {L}}_{Z^0\chi^+\chi^-}=\frac{g}{\cos{\theta_{w}}}Z_{\mu}
[\overline{\chi^+_i}\gamma^{\mu}(L'_{ij}P_L+R'_{ij}P_R)\chi^+_{j}]
\end{equation}
\begin{equation}
{\cal {L}}_{Z^0\chi^0\chi^0}=\frac{g}{\cos{\theta_{w}}}Z_{\mu}\frac{1}{2}
[\overline{\chi^0_i}\gamma^{\mu}(L''_{ij}P_L+R''_{ij}P_R)\chi^0_j]
\end{equation}
where
\begin{equation}
L'_{ij}=-V_{i1}V^*_{j1}-\frac{1}{2}V_{i2}V^*_{j2}+\delta_{ij}\sin^2{\theta_w}
\end{equation}
\begin{figure}
\hspace*{-0.6in}
\centering
\includegraphics[width=8cm,height=4cm]{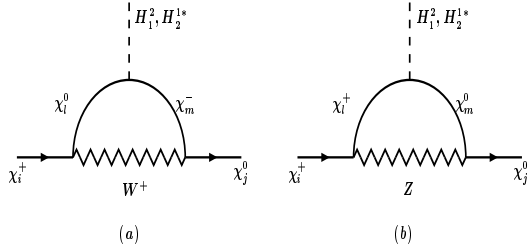}
\caption{The Chargino-Neutralino exchange contributions.}                                                                              		  	                              
\label{figef}
\end{figure}
\begin{equation}
R'_{ij}=-U^*_{i1}U_{ji}-\frac{1}{2}U^*_{i2}U_{j2}+\delta_{ij}\sin^2{\theta_w}
\end {equation}
\begin{equation}
L''_{ij}=-\frac{1}{2}X^*_{3i}X_{3j}+\frac{1}{2}X^*_{4i}X_{4j}
\end{equation}
\begin{equation}
R''_{ij}=\frac{1}{2}X_{3i}X^*_{3j}-\frac{1}{2}X_{4i}X^*_{4j}
\end{equation}
Using the above one finds 
\begin{equation}
\Delta\xi^{(3b)}_{ji}=-\sum_{l=1}^2\sum_{m=1}^4\frac{4g^2}{\cos^2{\theta_w}}
R''_{jm}L'_{li}\xi'_{ml}\frac{m_{\chi^+_l}m_{\chi^0_m}}
{16\pi^2}f(m^2_{\chi^+_l},m^2_{\chi^0_m},m^2_{Z^0})
\end{equation}
\noindent
{ Loop Fig. (4)}:
Here we need  the interactions of  $H^0_k\chi^+\chi^-$
and $H^0_k\chi^0\chi^0$
which are given by
\beqn
{\cal {L}}_{H^0_k\chi^+\chi^-}=-g\overline{\chi^+_i}[(Q^*_{ij}(Y_{k1}
-i Y_{k3}\sin\beta)+\nonumber\\
S^*_{ij}(Y_{k2}-iY_{k3}\cos\beta))P_L
+(Q_{ji}(Y_{k1}
+i Y_{k3}\sin\beta)+S_{ji}(Y_{k2}+iY_{k3}\cos\beta))P_R]
\chi^+_j H^0_k
\eeqn
\beqn
{\cal {L}}_{H^0_k\chi^0\chi^0}=-\frac{g}{\sqrt{2}}
\overline{\chi^0_i}[(Q^{'*}_{ij}(Y_{k1}
-i Y_{k3}\sin\beta)-S^{'*}_{ij}(Y_{k2}-iY_{k3}\cos\beta))
P_L\nonumber\\
+(Q^{'}_{ji}(Y_{k1}
+i Y_{k3}\sin\beta)-S^{'}_{ji}(Y_{k2}+iY_{k3}\cos\beta))P_R]
\chi^0_j H^0_k
\eeqn
Where
\begin{equation}
Q_{ij}=\frac{1}{\sqrt{2}}U_{i2} V_{j1},
S_{ij}=\frac{1}{\sqrt{2}}U_{i1} V_{j2}
\end{equation}
\begin{equation}
Q^{'}_{ij}=\frac{1}{\sqrt{2}}[X^*_{3i}(X^*_{2j}-\tan\theta_W X^*_{1j})],
S^{'}_{ij}=\frac{1}{\sqrt{2}}[X^*_{4j}(X^*_{2i}-\tan\theta_W X^*_{1i})]
\end{equation}
and the matrix elements $Y_{ij}$ are those of the diagonalizing matrix of
the  neutral  Higgs mass$^2$ matrix $M^2_{Higgs}$ such that
\begin{equation}
Y M^2_{Higgs} Y^T = diag(M^2_{H_1},M^2_{H_2},M^2_{H_3})
\end{equation}
 where in the limit of no CP violation 
$(H_1, H_2, H_3)\rightarrow (H^0, h^0, A)$ where $H^0$ ($h^0$) are the CP 
even heavy (light) neutral Higgs and $A$ is the CP odd Higgs. 
Using the product $P_L P_R =0$ we find that 
\begin{equation}
\Delta\xi^{(4)}=0
\end{equation}
\subsection{Loop analysis  of $\delta\xi_{ji}$ } 
For the loop corrections $\delta\xi_{ij}$ it is easy to see that 
\begin{equation}
\delta\xi^{(1b)}_{ji}=\delta\xi^{(2b)}_{ji}=\delta\xi^{(3a)}_{ji}=\delta\xi^{(3b)}_{ji}=0
\end{equation}
on using the properties  of the projection operators $P_L$ and  $P_R$  
, i.e., $P_LP_R=0$ and  $\gamma^{\mu}P_R=P_L\gamma^{\mu}$. 
Thus the only non vanishing $\delta\xi_{ji}$ are 
$\delta\xi_{ji}^{(1a)}$, $\delta\xi_{ji}^{(2a)}$
and $\delta\xi_{ji}^{(4)}$  and for these  
the computation following the same procedure as in Sec.(2.1) gives the 
following 
 \begin{equation}
\delta\xi^{(1a)}_{ji}=-\sum_{k=1}^2\sum_{n=1}^2\sqrt{2}gK_{t}V^*_{i2}D_{b1k}
\eta^*_{kn}
(\beta^*_{tj}D^*_{t1n}+\alpha_{tj}D^*_{t2n})\frac{m_t}{16\pi^2} 
f(m^2_t,m^2_{\tilde{b_k}},m^2_{\tilde{t_n}})
\end{equation}
\begin{equation}
\delta\xi^{(2a)}_{ji}=\sum_{k=1}^2\sum_{n=1}^2\sqrt{2}
g(V^*_{i1}D^*_{t1k}-K_tV^*_{i2}D^*_{t2k})\eta^*_{nk}(\alpha_{bj}D_{b1n}-\gamma_{bj}D_{b2n})
\frac{m_b}{16\pi^2}
f(m^2_b,m^2_{\tilde{t_k}},m^2_{\tilde{b_n}})
\end{equation}
\begin{equation}
\delta\xi^{(4)}_{ji}=\sum_{m=1}^4\sum_{l=1}^2\sum_{k=1}^3
\frac{g^2}{\sqrt{2}}\xi_{ml} B_{ilkmj}
\frac{m_{\chi^+_l}m_{\chi^0_m}}{16\pi^2}
f(m^2_{\chi^+_l},m^2_{\chi^0_m},m^2_{H^0_k})
\end{equation}
where
\beqn
B_{ilkmj}=[Q^{'*}_{jm}(Y_{k1}-iY_{k3}\sin\beta)-
S^{'*}_{jm}(Y_{k2}-iY_{k3}\cos\beta)]\nonumber\\
\times[Q^{*}_{li}(Y_{k1}-iY_{k3}\sin\beta)+
S^{*}_{li}(Y_{k2}-iY_{k3}\cos\beta]
\eeqn
\subsection{Loop analysis of $\Delta\xi_{ji}'$} 
Analogous to the analysis of Sec.(2.1) we may also decompose 
$\Delta\xi_{ji}'$ as follows corresponding to contributions arising
 from the loop diagrams  of Figs.(1)-(3) so that 
\beqn
\Delta\xi_{ji}'= \Delta\xi_{ji}^{'(1a)} + \Delta\xi_{ji}^{'(1b)} +
\Delta\xi_{ji}^{'(2a)} +\Delta\xi_{ji}^{'(2b)}+ 
\Delta\xi_{ji}^{'(3a)}+   \Delta\xi_{ji}^{'(3b)}
+\Delta\xi_{ji}^{'(4)}
\eeqn
Following the same procedure as in Sec.(2.1) we compute the 
contributions of various  and find the following results.
\begin{equation}
\Delta\xi'^{(1a)}_{ji}=\sum_{k=1}^2\sum_{n=1}^2\sqrt{2}g(U_{i1}D_{b1k}-K_bU_{i2}D_{b2k})(\eta^*_{kn})
(\alpha^*_{tj}D^*_{t1n}-\gamma^*_{tj}D^*_{t2n})\frac{m_t}{16\pi^2}
f(m^2_t,m^2_{\tilde{b_k}},m^2_{\tilde{t_n}})
\end{equation}
\begin{equation}
\Delta\xi'^{(1b)}_{ji}=-\sum_{k=1}^2g^2\frac{m_t}{m_W\sin{\beta}}
(K_bU_{i2}D^*_{t1k})
(\beta_{tj}D_{t1k}+\alpha^*_{tj}D_{t2k})\frac{m_tm_b}{16\pi^2}
f(m^2_t,m^2_b,m^2_{\tilde{t_k}})
\end{equation}
\begin{equation}
\Delta\xi'^{(2a)}_{ji}=-\sum_{n=1}^2\sum_{k=1}^2\sqrt{2}
g(K_bU_{i2}D^*_{t1k}) (\eta^*_{nk})
[\beta_{bj}D_{b1n}+\alpha^*_{bj}D_{b2n}]\frac{m_b}{16\pi^2}
f(m^2_b,m^2_{\tilde{t_k}},m^2_{\tilde{b_n}})
\end{equation}
\begin{equation}
\Delta\xi'^{(2b)}_{ji}=\sum_{k=1}^2g^2\frac{m_t}{m_W\sin{\beta}}
(U_{i1}D_{b1k}-K_bU_{i2}D_{b2k})
(\alpha^*_{bj}D^*_{b1k}-\gamma^*_{bj}D^*_{b2k})\frac{m_tm_b}{16\pi^2}
f(m^2_b,m^2_t,m^2_{\tilde{b_k}})
\end{equation}
\begin{equation}
\Delta\xi'^{(3a)}_{ji}=-\sum_{m=1}^2\sum_{l=1}^44g^2L^*_{jm}R_{li}\xi_{lm}
\frac{m_{\chi^0_l}m_{\chi^-_{m}}}{16\pi^2}
f(m^2_{\chi^0_l},m^2_{\chi^-_m},m^2_W)
\end{equation}
\begin{equation}
\Delta\xi'^{(3b)}_{ji}=-\sum_{l=1}^2\sum_{m=1}^4\frac{4g^2}
{\cos^2{\theta_W}}L''^*_{jm}R'_{li}\xi_{ml}
\frac{m_{\chi^+_l}m_{\chi^0_m}}{16\pi^2}f(m^2_{\chi^+_l},m^2_{\chi^0_m},m^2_Z)
\end{equation}
\begin{equation}
\Delta\xi'^{(4)}=0
\end{equation}
\subsection{Loop analysis of $\delta\xi'_{ji}$}  
An analysis similar to that of Sec.(2.3) gives 
\begin{equation}
\delta\xi'^{(1b)}_{ji}=\delta\xi'^{(2b)}_{ji}=\delta\xi'^{(3a)}_{ji}=
\delta\xi'^{(3b)}_{ji}=0
\end{equation}
and the only non vanishing elements are 
\begin{equation}
\delta\xi'^{(1a)}_{ji}=\sum_{k=1}^2\sum_{n=1}^2\sqrt{2}g(U_{i1}D_{b1k}-
K_bU_{i2} D_{b2k})
\eta'_{kn}
(\alpha^*_{tj}D^*_{t1n}-\gamma^*_{tj}D^*_{t2n}) \frac{m_t}{16\pi^2
}f(m^2_t,m^2_{\tilde{b_k}},m^2_{\tilde{t_n}}),
\end{equation}

\begin{figure}
\hspace*{-0.6in}
\centering
\includegraphics[width=4cm,height=4cm]{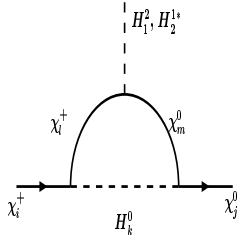}
\caption{Higgs exchange contributions.}                                                                              		  	                              
\label{figh}
\end{figure}
\begin{equation}
\delta\xi'^{(2a)}_{ji}=-\sum_{k=1}^2\sum_{n=1}^2\sqrt{2}gK_bU_{i2}D^*_{t1k}
(\eta'_{nk})(\beta_{bj}D_{b1n} + \alpha^*_{bj}D_{b2n})
\frac{m_b}{16\pi^2}f(m^2_b,m^2_{\tilde{t_k}},m^2_{\tilde{b_n}})
\end{equation}
and
\begin{equation}
\delta\xi'^{(4)}_{ji}=\sum_{m=1}^4\sum_{l=1}^2\sum_{k=1}^3
\frac{g^2}{\sqrt{2}}\xi^{'}_{ml} A_{ilkmj}
\frac{m_{\chi^+_l}m_{\chi^0_m}}{16\pi^2}
f(m^2_{\chi^+_l},m^2_{\chi^0_m},m^2_{H^0_k})
\end{equation}
where
\beqn
A_{ilkmj}=[Q^{'}_{mj}(Y_{k1}+iY_{k3}\sin\beta)-
S^{'}_{mj}(Y_{k2}+iY_{k3}\cos\beta)]\nonumber\\
\times[Q_{il}(Y_{k1}+iY_{k3}\sin\beta)+
S_{il}(Y_{k2}+iY_{k3}\cos\beta]
\eeqn
\subsection{Charged Higgs Decays Including Loop Effects} 
We summarize now the result of the analysis. 
Thus $L_{eff}$ of $Eq.(6)$ may be written as follows
\begin{equation}
{\cal {L}}_{eff}=H^-\overline{\chi^0_j}(\alpha^S_{ji}+\gamma_{5}\alpha^P_{ji})\chi^+_i+ 
H.c
\end{equation}
where
\begin{equation}
\alpha^{S}_{ji}=\frac{1}{2}(\xi'_{ji}+\delta\xi'_{ji})\sin{\beta}+
\frac{1}{2}\Delta\xi'_{ji}\cos{\beta}+\frac{1}{2}(\xi_{ji}+
\delta\xi_{ji})\cos{\beta}+\frac{1}{2}\Delta\xi_{ji}\sin{\beta}
\label{scoupling}
\end{equation}
and where
\begin{equation}
\alpha^{P}_{ji}=\frac{1}{2}(\xi'_{ji}+\delta\xi'_{ji})\sin{\beta}
+\frac{1}{2}\Delta\xi'_{ji}\cos{\beta}-
\frac{1}{2}(\xi_{ji}+\delta\xi_{ji})\cos{\beta}-
\frac{1}{2}\Delta\xi_{ji}\sin{\beta}
\label{pcoupling}
\end{equation}
Next we discuss the implications of the above result for the 
decay of the charged Higgs.
The effect of loop corrections on the charged Higgs decays into
$\bar t b$ ($t\bar b$) and into $\tau^-\bar\nu_{\tau}$ 
($\tau^+\nu_{\tau}$) was exhibited in the analysis of 
Ref.\cite{Ibrahim:2003tq} but charged Higgs decays into $\chi^{\pm}_i\chi^0_j$
were not taken into account. However, if the kinematics allows 
the decay of $H^{\pm}$ into $\chi^{\pm}_i\chi_j^0$
then all the allowed modes must be included and the 
analysis of Ref.\cite{Ibrahim:2003tq} along with the analysis given here 
allows one to do an analysis including one loop corrections
of the branching ratios. We note in passing that the CP phases
enter in the effective couplings and thus branching ratios
will be sensitive to the CP phases. Specifically in the 
analysis given in this section the CP phases enter via the 
diagonalizing matrices U and V from the chargino sector,
via the matrix X in  the neutralino sector and via the matrix Y
in the Higgs sector.
 Before proceeding
further we give below the decay widths in terms of the 
effective couplings of Eq.~(\ref{scoupling}) and of 
Eq.~(\ref{pcoupling}). One has for the decay of 
$H^-$ into $\chi^0_j\chi^-_i$ (j=1,2; i=1,2,3,4) the result
\beqn
\Gamma_{ji}(H^{-}\rightarrow\chi^0_j\chi^-_i)=\frac{1}{4\pi 
M^3_{H^-}}
\sqrt{[(m^2_{\chi^0_j}+m^2_{\chi^{+}_i}-M^2_{H^{-}})^2
-4m^2_{\chi^{+}_i}m^2_{\chi^0_j}]}\nonumber\\
([\frac{1}{2}((|\alpha^{S}_{ji}|)^2+(|\alpha^{P}_{ji}|)^2)
(M^2_{H^{-}}-m^2_{\chi^{+}_i}-m^2_{\chi^{0}_j})
-\frac{1}{2}((|\alpha^{S}_{ji}|)^2-(|\alpha^{P}_{ji}|)^2)
(2m_{\chi^{+}_i}m_{\chi^{0}_j})])
\label{branching}
\eeqn
The analysis of this section is utilized in Sec.(3) where we 
give a numerical analysis of the size of the loop effects and
discuss the effect of the loop corrections on the branching 
ratios.
\begin{table}[t]
\begin{center}
\begin{tabular}{|l|l|l|}
\hline
 $|d_e| e.cm$ &  $|d_n| e.cm$ &  $C_{Hg} cm$  \\ \hline
 $2.69 \times 10^{-27}$ & $3.37 \times 10^{-26}$ & $2.15 \times 10^{-26}$ 
\\ \hline 
\end{tabular}
\end{center}
\caption{The EDMs for the case when  
$m_A=950$, $m_0=275$,  $m_{\frac{1}{2}}=270$,
$\xi_1=.59$,  $\xi_2=.65$,  $\xi_3=.655$,  $\alpha_{A_0}=1.0$,  $A_0=4$, 
$\theta_{\mu}=2.5$, and  $\tan\beta=50$.  All masses are in GeV and 
all angles are in radians. $C_{Hg}$ is as defined in Ref.\cite{olive}. }  
\label{edm_value}
\end{table}
\section{Numerical Analysis}
\begin{figure}           
\vspace*{-2.0in}                                 
\subfigure[]{                       
\label{perc_params_a} 
\hspace*{-0.6in}                     
\begin{minipage}[b]{0.4\textwidth}                       
\centering
\includegraphics[width=\textwidth,height=\textwidth]{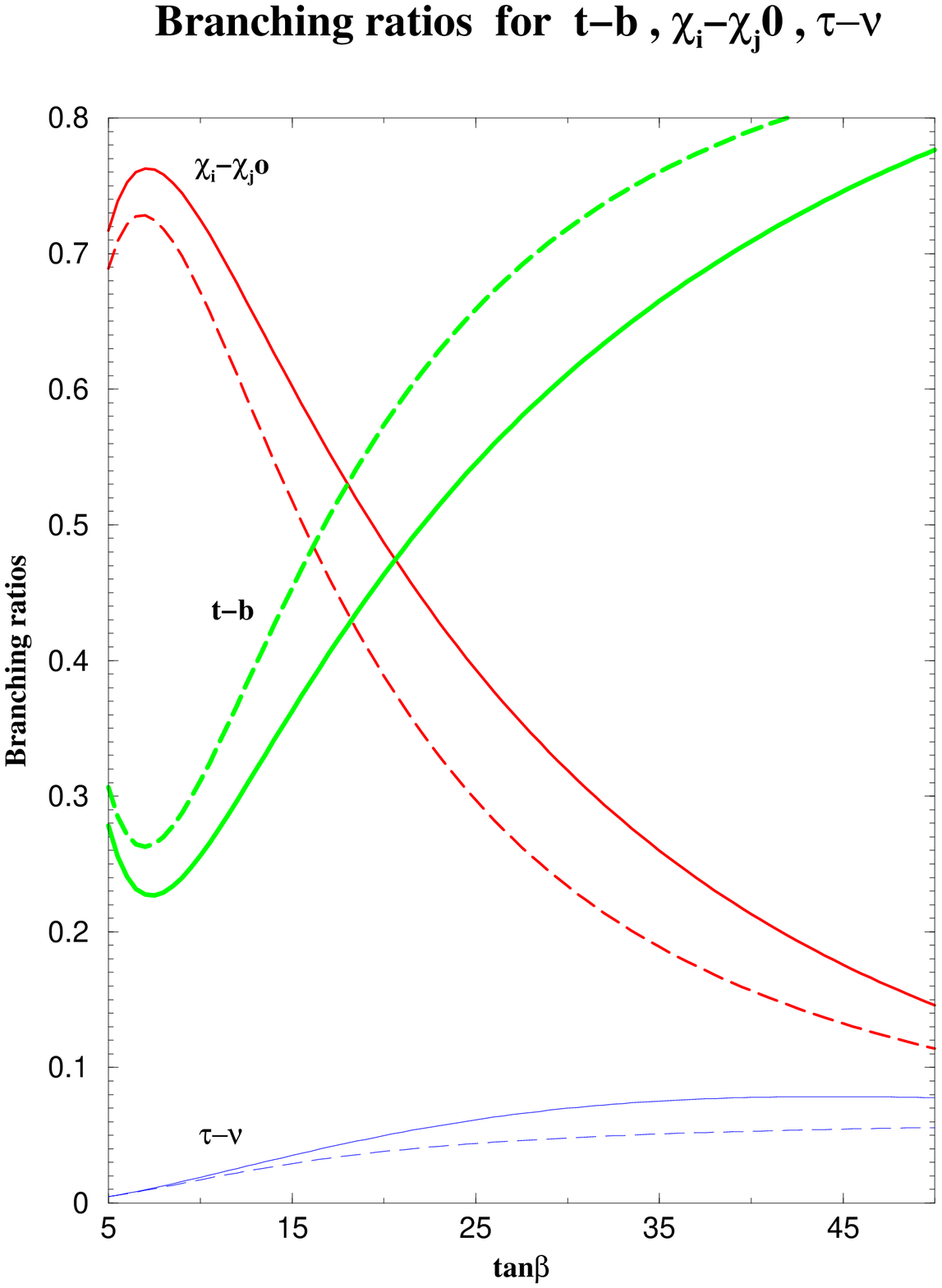}    
\end{minipage}}                       
\hspace*{0.3in}
\subfigure[]{                       
\label{perc_params_b}                       
\begin{minipage}[b]{0.4\textwidth}                       
\centering                      
\includegraphics[width=\textwidth,height=\textwidth]{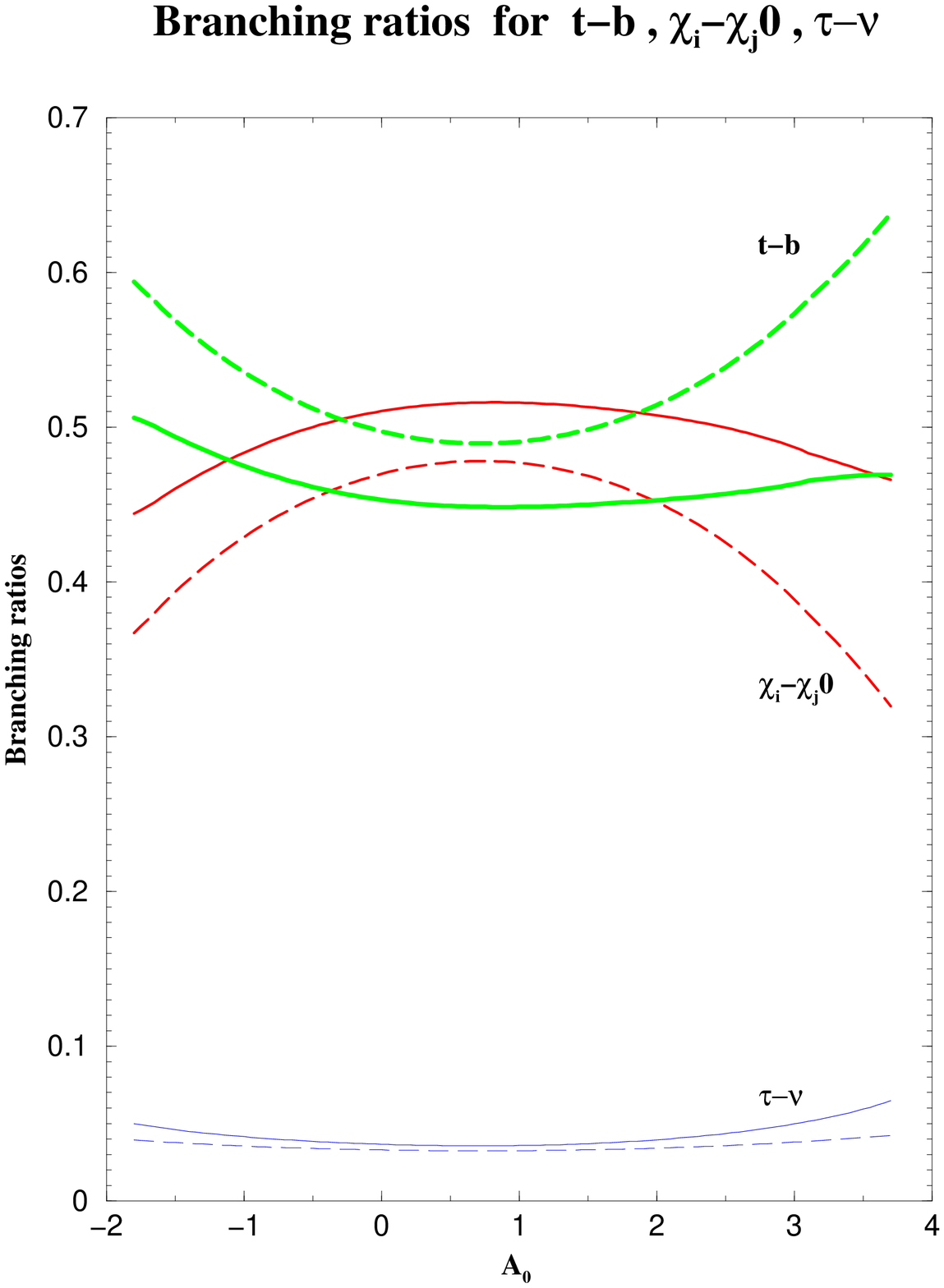} 
\end{minipage}}                       
\hspace*{-0.6in}                     
\subfigure[]{                       
\label{perc_params_c}                      
\begin{minipage}[b]{0.4\textwidth}                       
\centering
\includegraphics[width=\textwidth,height=\textwidth]{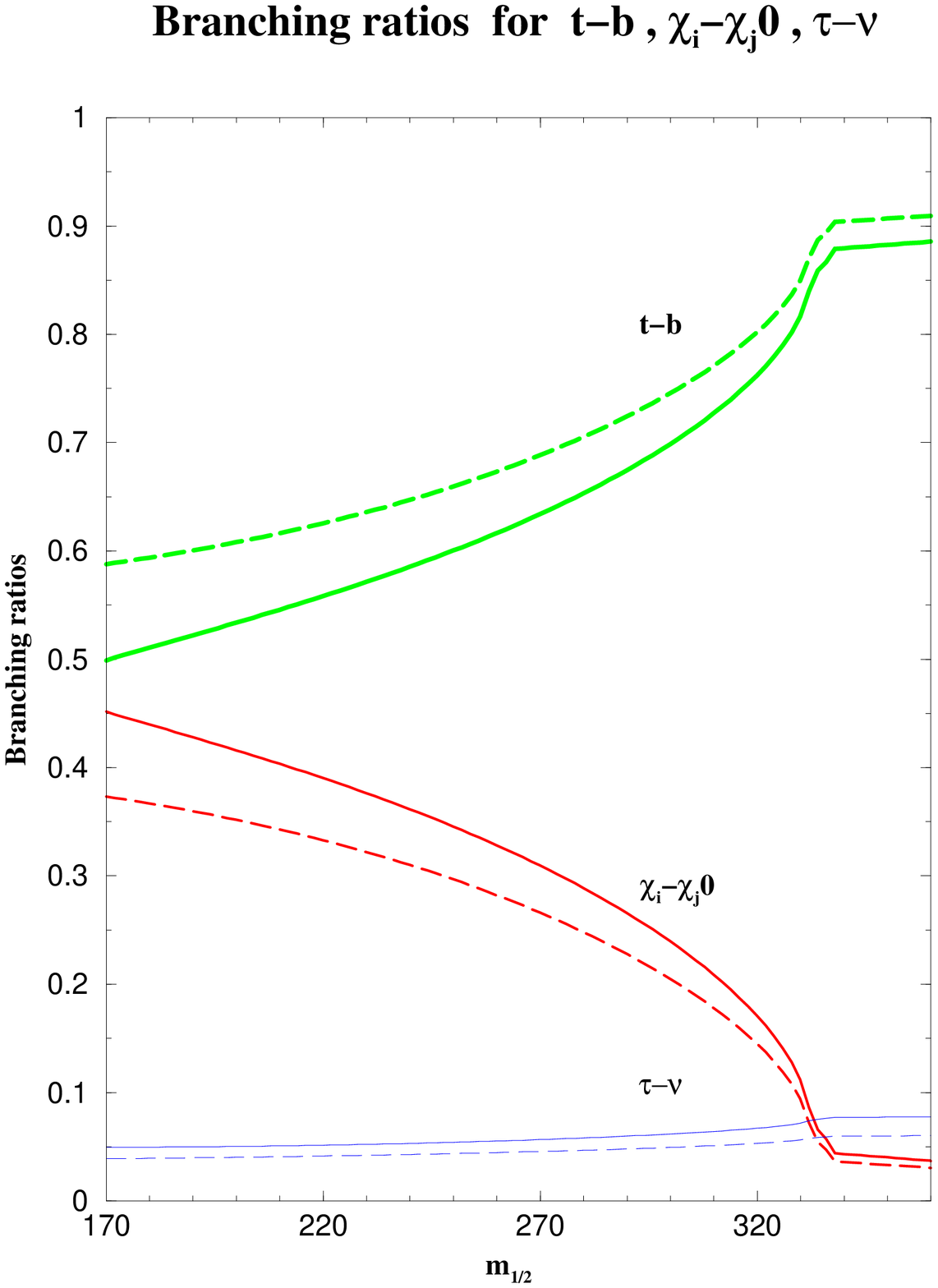}
\end{minipage}}
\hspace*{0.3in}                       
\subfigure[]{                       
\label{perc_params_d}                       
\begin{minipage}[b]{0.4\textwidth}                       
\centering                      
\includegraphics[width=\textwidth,height=\textwidth]{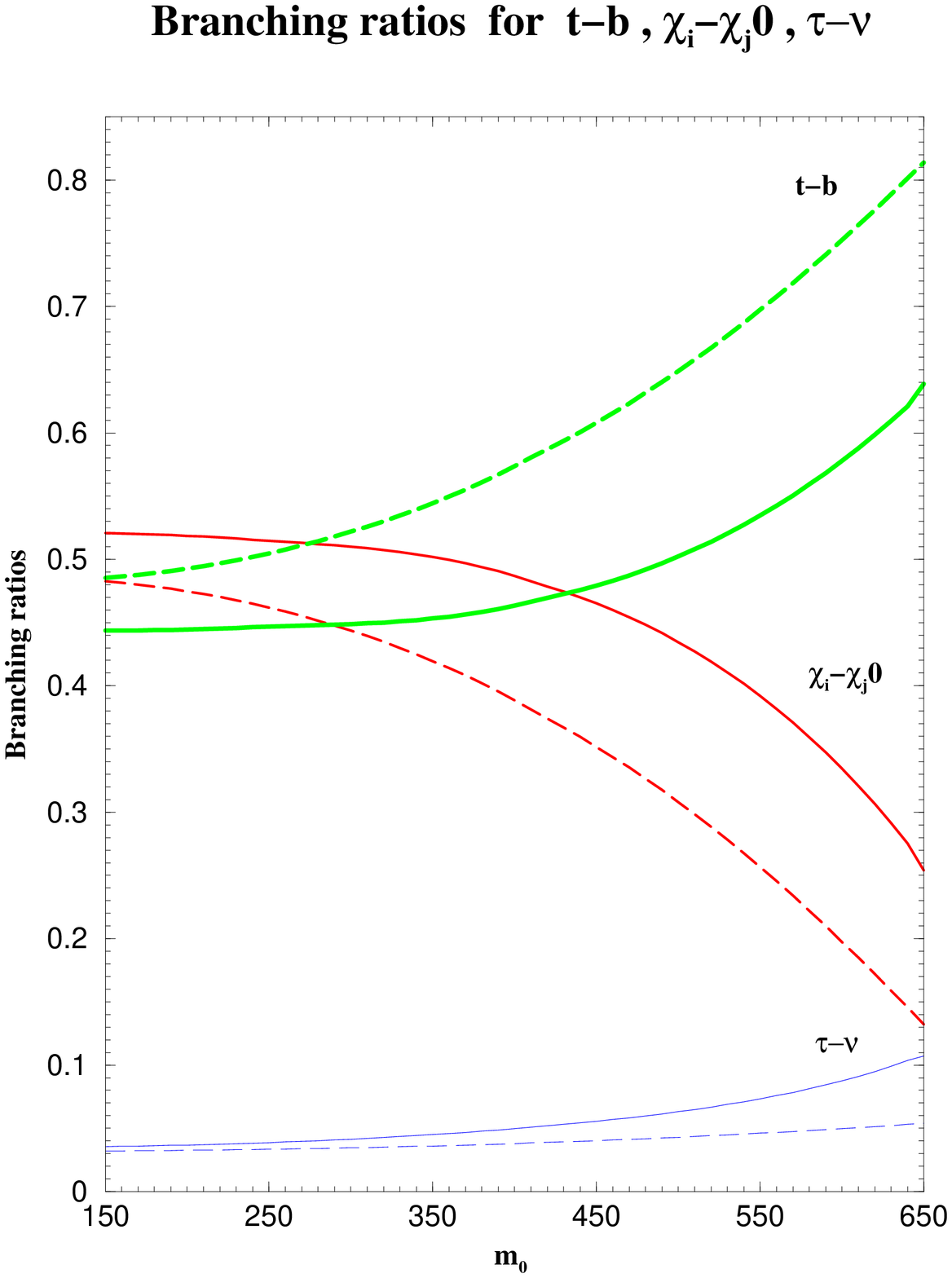}
\end{minipage}}                       
\caption{
Plot of branching ratios for the decay of $H^{\pm}$ as a function of 
$\tan\beta$ in (a), as a function of $A_0$ in (b), as a function of 
$m_{\frac{1}{2}}$ in (c) and as a function of $m_0$  in (d). 
The  parameters are 
$m_A=800$, $m_0=400$, $m_{\frac{1}{2}}=140$, $A_0$=3, $\tan\beta =20$, 
 $\xi_1=0$,  $\xi_2=0$,  $\xi_3=0$,  $\theta_{\mu}=0$,
 $\alpha_{A_0}=0$ except that the running parameter is to be deleted from the
 set for a given subgraph. The long dashed lines are the branching
ratios at the tree level while the solid lines include the 
loop correction. The curves labelled $\chi_i^-\chi_j^0$ here and in
Fig.(6) stand for sum of branching ratios into all allowed 
$\chi_i^-\chi_j^0$ modes. All masses are in unit of GeV and 
all angles in unit of radian.}                                                               
\label{perc_params} 
\end{figure}

The analysis of loop corrections given in Sec.(2) is quite
general as they are computed within the framework of MSSM.
However, the parameter space of MSSM is rather large, and for the
purpose of numerical computations it is desirable to restrict
the analysis to a more constrained space. Here we will
use the framework of the extended SUGRA model for this purpose.
Thus we assume the parameter space of the model to consist
of  $m_A$ (mass of the  CP odd Higgs boson), 
 $\tan\beta$, complex trilinear coupling 
$A_0$,   $SU(3)$, $SU(2)$ and $U(1)_Y$ gaugino masses
$\tilde m_i =m_{\frac{1}{2}} e^{i\xi_i}$ (i=1,2,3) and $\theta_{\mu}$, where
$\theta_{\mu}$ is the phase of $\mu$. The analysis is carried out
by evolving the soft parameters from the grand unification scale
to the electroweak scale and $ |\mu|$
is determined by radiative breaking of the electroweak symmetry
(see, for example, Ref.\cite{renorm})
while $\theta_{\mu}$ remains an arbitrary parameter. We note in 
passing that not all the phases are arbitrary as only 
specific combinations of the phases appear in the determination
of physical quantities\cite{inmssm}. 
 We discuss now the size of the loop correction to the branching ratios.
 Typically in the parameter space investigated the squarks and the 
  sleptons  are too heavy to be produced as final states in the
  decay of the charged Higgs. Further, 
the decay modes  $H^{\pm}\rightarrow W^{\pm} H^0_k$ contribute
 less than 1$\%$ to the total Higgs decay  
due to a mixing angle suppression factor{\cite{Gunion:1988yc}}.
The decay modes of charged higgs into quarks and leptons of the first and
second families can be safely ignored compared to the contribution 
of the third family due to the smallness of the Yukawa couplings 
of the first two families.
Thus the decay of the charged Higgs is dominated  by 
 the following modes: top-bottom, chargino-neutralino
and tau-neutrino. 
In Fig.~\ref{perc_params_a}
 we give a plot of the branching ratios
of $H^-$ to $\bar t b$, $\tau^- \bar \nu_{\tau}$ and 
$\chi_i^-\chi_j^0$ as a function of $\tan\beta$. The  analysis is
given at the tree level and also including the loop correction.
One finds the loop correction to be substantial reaching $20\%$ 
or more in a  significant part of the parameter space. 
We  note that the chargino branching ratio is substantial and
for small $\tan\beta$ the dominant one.
We also note that  the branching ratio for  $\bar t b$ 
at the tree level  exhibits a minimum at
$\tan\beta \simeq 7$ while the branching ratio for 
$\chi_i^-\chi_j^0$  at the tree level exhibits a 
maximum at almost the same value. 
Further, the position of these extrema are essentially left
intact when one includes the loop correction. 

To understand the above phenomena we need to consider the partial 
width expression for
the various decay modes. Thus at the 
 tree level the partial width for the decay mode  $\bar t b$  may 
 be expressed as
\beq
\Gamma^{tree}_{\bar t b}=\alpha_1 (m_b^2 \tan^2\beta +m_t^2 \cot^2\beta) +\alpha_2
\eeq
where $\alpha_{1,2}$ are functions of the masses and the couplings but are
independent of $\tan\beta$. Clearly then $\Gamma^{tree}_{\bar t b}$
has a minimum at 
\beq
\tan\beta=\sqrt{\frac{m_t}{m_b}}
\label{tanbeta}
\eeq
Similarly the decay width in the chargino-neutralino channel may be  be expressed as
\beq
\Gamma^{tree}_{\chi_i^-\chi_j^0}=g^2 [\sin^2\beta f_1(X,U,V,M)+\sin\beta \cos
\beta f_2(X,U,V,M)]
\eeq
where $f_{1,2}$ are functions of the matrix  X which diagonalizes
the neutralino mass
matrix, and  of matrices U and V  which diagonalize the chargino mass matrix. 
They are also functions of the eigen spectrum  of 
the charged Higgs, chargino and neutralino mass matrices. Now the matrices
X, U and V and the eigen spectrum of 
the chargino-neutralino mass matrices  are functions of 
$\tan\beta$ and thus $\Gamma^{tree}_{\chi_i^-\chi_j^0}$ 
are complicated functions of
$\tan\beta$. However, numerical studies of these functions show that they are weak
functions of $\tan\beta$.
Finally the decay width in the $\bar{\nu}\tau$ channel may be written as
\beq
\Gamma^{tree}_{\bar{\nu} \tau}=\alpha_3 \tan^2\beta
\eeq
where $\alpha_3$ is a function of the masses and couplings but is
independent of $\tan\beta$.
The loop correction to different decay widths 
 is generally different. In the $\bar t b$
channel the contribution of the loop correction to Yukawa couplings $\overline{\Delta h}_b$, 
$\overline{\delta h}_b$, $\overline{\Delta h}_t$
and $\overline{\delta h}_t$\cite{Ibrahim:2003tq}
  reduce  $\Gamma_{\bar t b}$ and the magnitude of this  reduction generally 
increases as $\tan\beta$ increases. Thus we find that the 
 branching ratio including the loop correction has the 
same behavior as the one at the tree level with a small separation between them
for small $\tan\beta$ and this separation tends to get larger as 
$\tan\beta$ increases.
Combined with the fact that the tree level minimum occurs at a
small value of $\tan\beta$, one finds that the inclusion of the
loop correction induces only a negligible displacement of the
minimum. 
 In the $\chi_i^-\chi_j^0$ channel, the loop effects come into play via 
the quantities $\Delta \xi_{ji}$, $\delta \xi_{ji}$, $\Delta \xi'_{ji}$
and $\delta \xi'_{ji}$. 
However, the effect of $\tan\beta$ on $\Gamma_{\chi^-_i\chi_j^0}$ of the 
chargino-neutralino  channel is still small even after considering the loop 
effects and since the top-bottom 
and chargino-neutralino modes are the largest we find that the branching ratio
for the chargino-neutralino channel has maxima almost at the same position 
as the minima for the  $\bar t b$ mode. Finally, the decay width for the  
$\tau^- \bar \nu_{\tau}$ mode
   increases as $\tan\beta$ increases  both at 
the tree and at the loop level.
The loop effects appear via the quantities 
$\overline{\Delta h}_{\tau}$ and  $\overline{\delta h}_{\tau}$.
 These corrections also lead to a $\tan^2\beta$ dependence
 of the loop corrected partial width for the $\tau\bar \nu_{\tau}$
 mode as seen in Fig.~\ref{perc_params_a}.
 
A similar analysis but as a function of $A_0$ is given in 
Fig.~\ref{perc_params_b}. Again the branching ratios into $\bar t b$ and 
into $\chi_i^-\chi_j^0$ are the largest and the loop correction
is again sizable reaching in this case as much as 25\% or more
for large values of $A_0$.
 At the tree level  $\Gamma^{tree}_{\bar t b}$ and
 $\Gamma^{tree}_{\tau^- \bar \nu_{\tau}}$ are indeed independent of $A_f$.
However $\Gamma^{tree}_{\chi_i^-\chi_j^0}$ is a function of $A_0$ since the value
of $\mu$ that enters the chargino and neutralino mass matrices depends on 
$A_0$ through the renormalization group evolution. Inclusion of the loop effects 
introduce additional contributions which are $A_0$ dependent
 through the matrix elements of the  diagonalizing matrices  D, Y, U, V and X. 
 The change of the partial width of the chargino-neutralino channel as 
$|A_0|$ changes is reflected  in the branching ratio analysis at both the
tree and the loop level as shown in  Fig.~\ref{perc_params_b}.
 
In Fig.~\ref{perc_params_c} we give an analysis of the branching ratios 
as a function of $m_{1/2}$.
The dependence of the branching ratio on $m_{\frac{1}{2}}$ is easily explained by
noting  that the tree level expressions
for the partial width of the  $\bar t b$ and the $\bar \nu_{\tau} \tau$ modes are independent
of $m_{1/2}$. However, the tree partial width for the chargino-neutralino 
mode decreases as  $m_{1/2}$ increases because of the kinematic supression.
In fact there is a kinematic cutoff beyond which this mode is not allowed.
So the effect of  $m_{1/2}$ on the tree level branching ratios comes directly from the
effect of this parameter on the chargino-neutralino mode. 
Inclusion of  the loop correction supresses the partial width of the  $\bar t b$
 mode and the magnitude of 
this suppression decreases as $m_{1/2}$ increases. The kinematic
suppression in the case of the chargino-neutralino mode still works as for 
 the tree level case. Using the combined effects of the above factors
 one finds that  the loop correction for the branching ratios
are largest for the smallest allowed values of $m_{1/2}$ and
become relatively smaller as $m_{1/2}$ becomes relatively larger.
This phenomenon is uniform between 
the three branching ratios plotted in Fig.~\ref{perc_params_c}.
Finally we note  that the sharp bend in the curves at the high end of 
 $m_{\frac{1}{2}}$ arises from  closing of 
 some of the chargino-neutralino modes  because the corresponding
 $\Gamma_{ij}$  vanish for those modes whose threshold 
 $(m_{\chi_i^{\pm}}+ m_{\chi_j^{0}})>m_{H^{\pm}}$.

\begin{figure}                       
\vspace*{-1.0in}                                 
\subfigure[]{
\label{figy9b}
\hspace*{-0.6in}                     
\begin{minipage}[b]{0.4\textwidth}                       
\centering
\includegraphics[width=\textwidth]{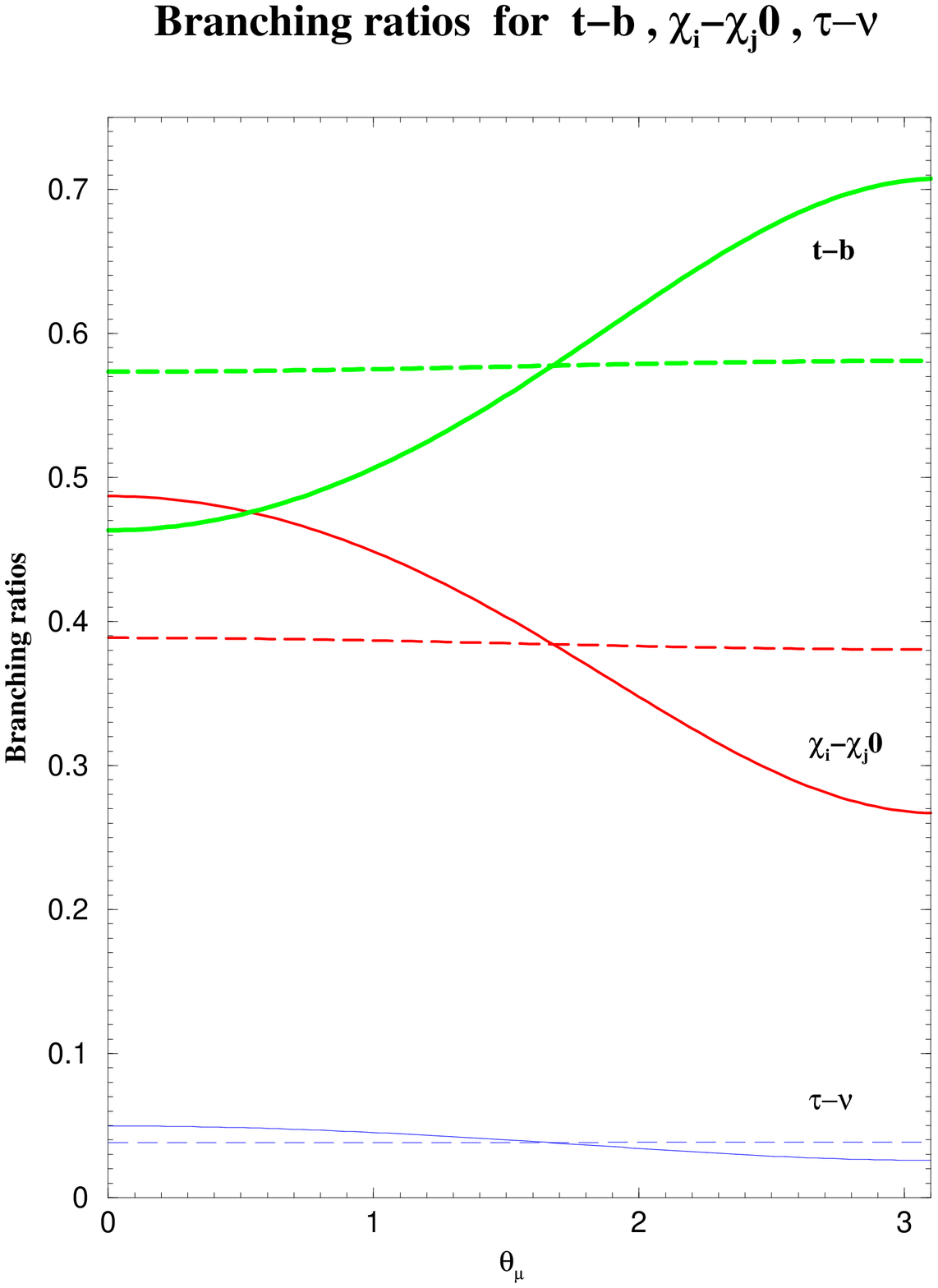}                      
\end{minipage}}
\hspace*{0.3in}                       
\subfigure[]{
\label{figy10b}    
\begin{minipage}[b]{0.4\textwidth}                       
\centering                      
\includegraphics[width=\textwidth]{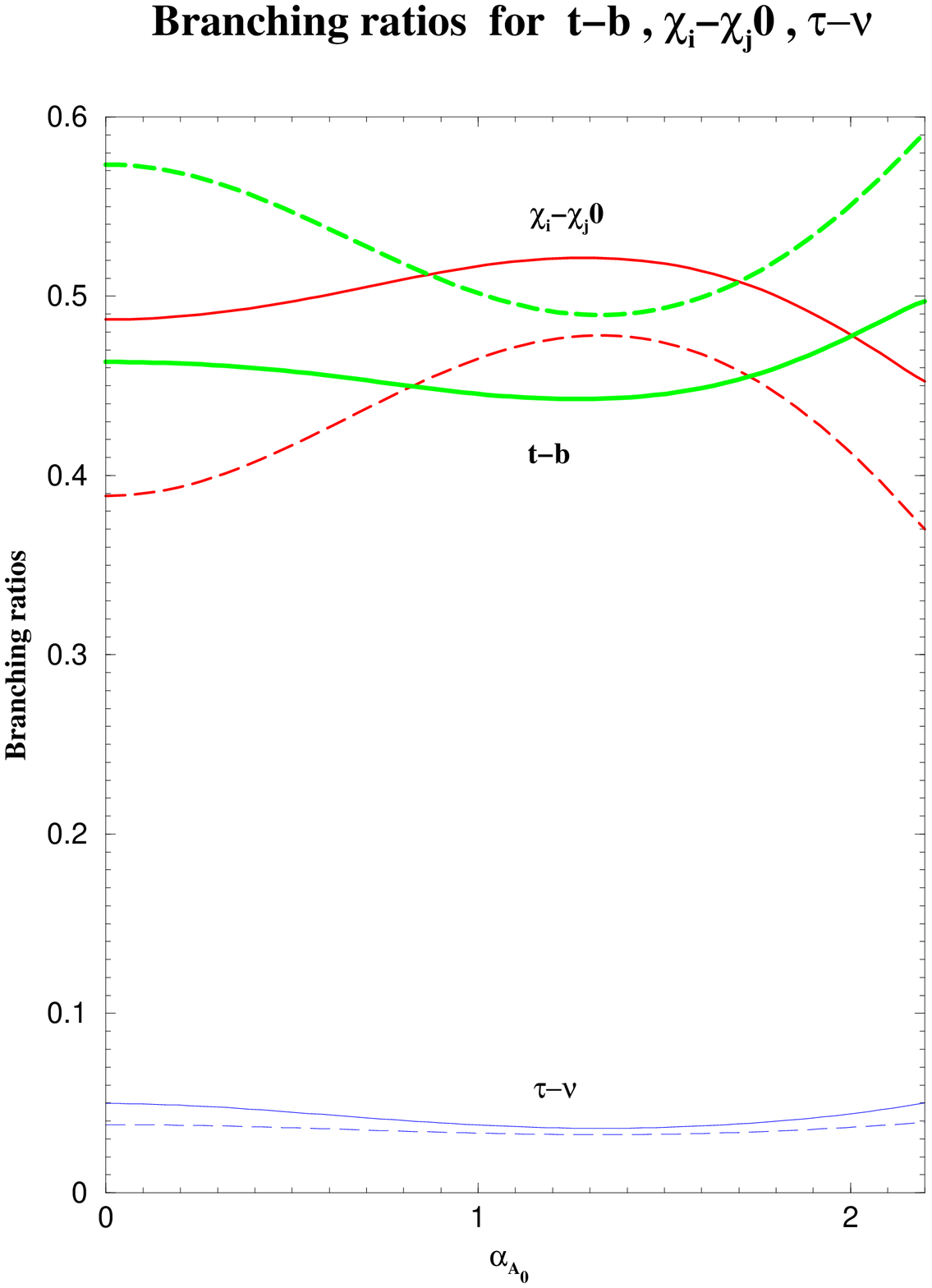}                      
\end{minipage}}                       

\subfigure[]{
\label{figz15}             
\hspace*{-0.4in}                               
\begin{minipage}[b]{\textwidth}                       
\centering                      
\includegraphics[width=0.4\textwidth]{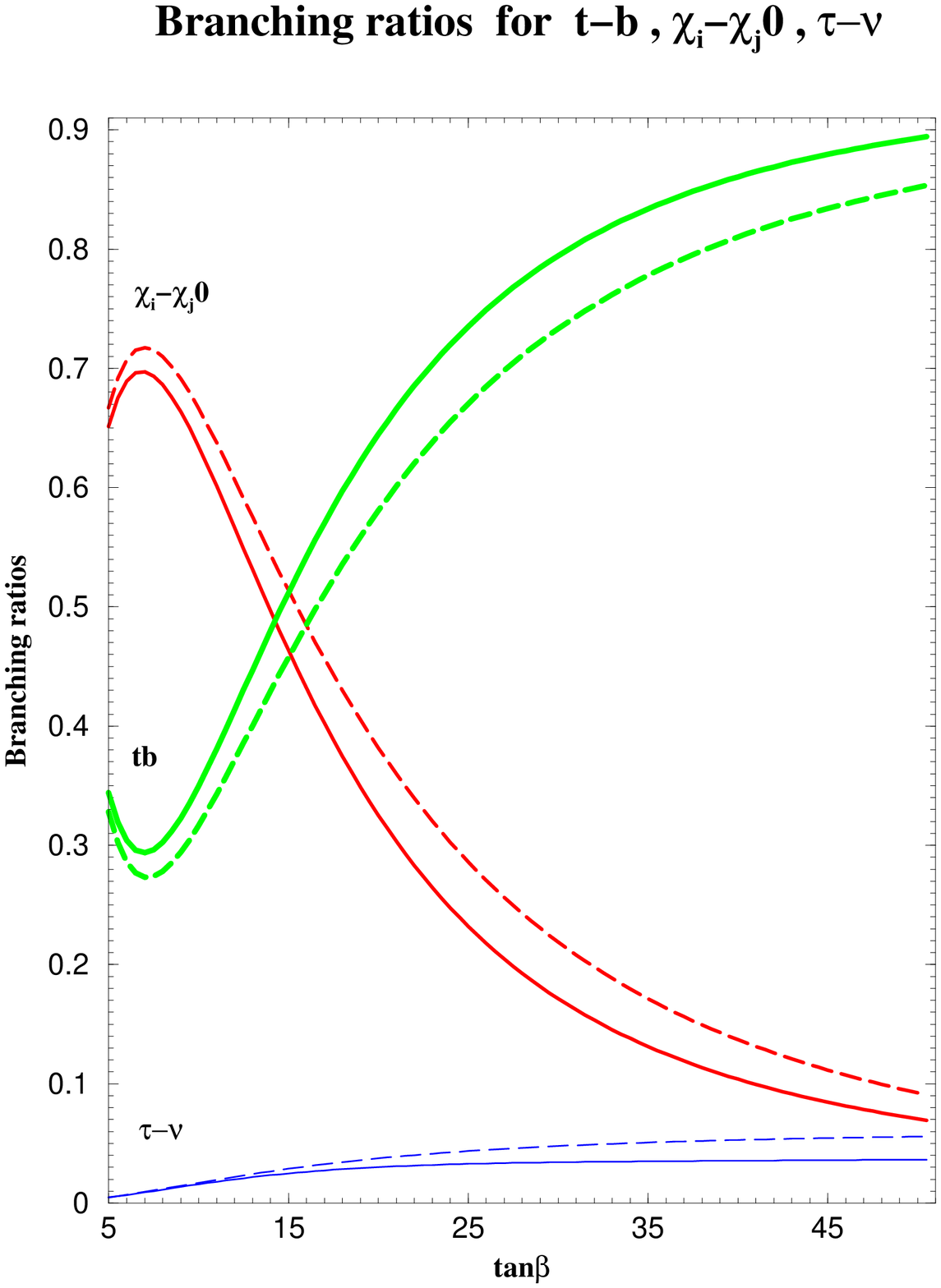}      
\end{minipage}}                       
\caption{
Plot of branching ratios for the decay of $H^{\pm}$ as a function of 
$\theta_{\mu}$ in (a) and as a function of $\alpha_{A_0}$ in (b).
 The  parameters are 
$m_A=800$, $m_0=400$, $m_{\frac{1}{2}}=140$, $A_0$=3, $\tan\beta =20$, 
 $\xi_1=0$,  $\xi_2=0$,  $\xi_3=0$,  $\theta_{\mu}=0$,
 $\alpha_{A_0}=0$ except that the running parameter is to be deleted from the
 set for a given subgraph.  The analysis of (c) corresponds to the 
 input of Table 1 except that $\tan\beta$ is a running parameter.
 The long dashed lines are the branching
ratios at the tree level while the solid lines include the 
loop correction. All masses are in unit of GeV and all angles in 
unit of radian.}                       
\label{unification_spectra}
\end{figure}

A similar analysis but as a function of  the universal scalar mass
$m_0$ is given in Fig.~\ref{perc_params_d}. Quite interestingly here
the loop correction gets larger as $m_0$ increases.
The difference between the behavior of the branching ratio as a function
of $m_0$ and as a function of $m_{1/2}$ after inclusion of the loop effects,
comes mainly 
from the fact that the loop corrected decay width for $\bar t b$ 
gets suppressed relative to its tree value  as $m_0$ increases.
This arises due to the fact that the mass splitting between the
squark mass eigen states that enter in the  $\bar t b$ decay mode
 increases because the trilinear coupling $m_0A_f$  increases as $m_0$
increases. Using the same reasoning 
one can explain the  splitting between 
the tree and loop corrected branching ratios  in Fig.~\ref{perc_params_b}.
Returning to Fig.~\ref{perc_params_d} we note that the loop correction
in  Fig.~\ref{perc_params_d} lies in the range of 10-30\%. 
In Fig.~\ref{figy9b}  we give a plot of the branching ratios 
with and without the loop correction as a function of the CP phase
$\theta_{\mu}$. 
At the tree level the branching ratios are flat as a function of 
$\theta_{\mu}$ since there is no dependence at the tree level of the
decay widths into $\bar tb$ and into $\tau\bar \nu$ 
on $\theta_{\mu}$ and  further in
part of the parameter space investigated the 
decay width into chargino-neutralino modes depends
only weakly on $\theta_{\mu}$. This situation changes
dramatically when the loop correction is included.
Thus the   inclusion of the loop correction
brings in a significant dependence on $\theta_{\mu}$. This arises 
mainly due to the 
  $\theta_{\mu}$ dependence of QCD  correction for the top-bottom
mode where there is gluino running in the loop that contributes to the
charged Higgs-top-bottom coupling.
A similar analysis as a function of the 
CP phase $\alpha_{A_0}$ is given in Fig.~\ref{figy10b}.
The analysis of the effect of  $\alpha_{A_0}$ is similar to 
the effect of $|A_0|$ on the branching ratios  as may be seen by a 
comparison of Fig.~\ref{perc_params_b} and  Fig.~\ref{figy10b}.
In Fig.~\ref{figz15} we exhibit the
results where the electric dipole moment (edm) constraints are included as given in Table 1. 
These analyses indicate that the loop correction is a sensitive 
function of  CP phases. 

\begin{figure}           
\vspace*{-2.0in}                                 
\subfigure[]{                       
\label{fig17a} 
\hspace*{-0.6in}                     
\begin{minipage}[b]{0.4\textwidth}                       
\centering
\includegraphics[width=\textwidth,height=\textwidth]{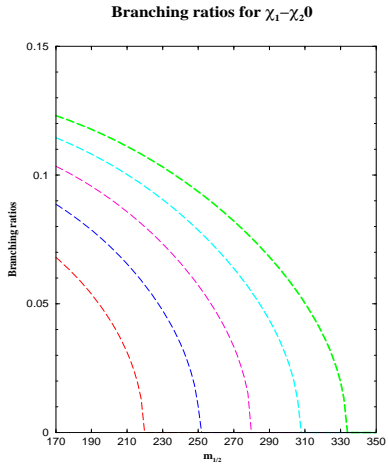}    
\end{minipage}}                       
\hspace*{0.3in}
\subfigure[]{                       
\label{fig17b}                        
\begin{minipage}[b]{0.4\textwidth}                       
\centering                      
\includegraphics[width=\textwidth,height=\textwidth]{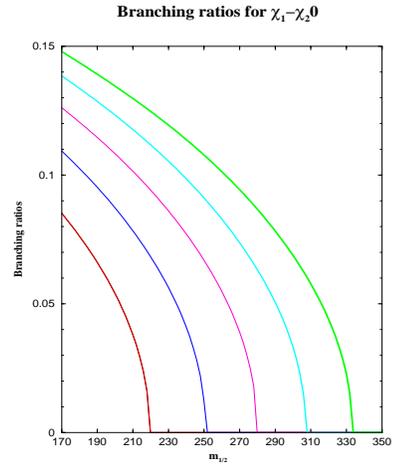} 
\end{minipage}}                       
\hspace*{-0.6in}                     
\subfigure[]{                       
\label{fig17c}                       
\begin{minipage}[b]{0.4\textwidth}                       
\centering
\includegraphics[width=\textwidth,height=\textwidth]{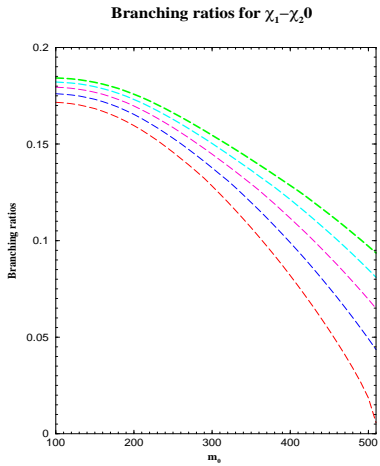}
\end{minipage}}
\hspace*{0.3in}                       
\subfigure[]{                       
\label{fig17d}                        
\begin{minipage}[b]{0.4\textwidth}                       
\centering                      
\includegraphics[width=\textwidth,height=\textwidth]{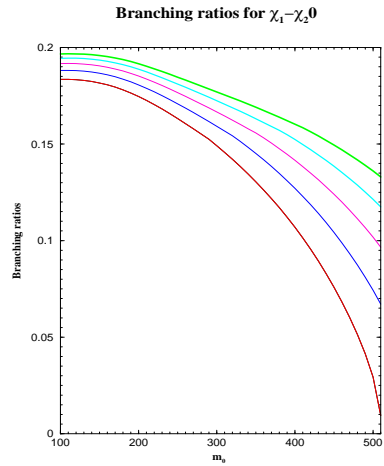}
\end{minipage}}                       
\caption{\small Plots of the branching ratio $\chi_1^{\pm}\chi_2^0$ 
  for the decay of $H^{\pm}$ as a function 
of $m_{\frac{1}{2}}$ ( (a)-(b)), and as a function of $m_{0}$ 
((c)-(d)). The common inputs are $\tan\beta=20$, $\xi_1=0$, $\xi_2=0$,  
$\xi_3=0$ , $A_0=3$, $\alpha_{A_0}=0$, and $\theta_{\mu}=0$
and $m_A$ ranges from $600-800$ in increments of $50$ in 
ascending order of curves.
(a) -(b) have the additional input $m_0=400$
while (c)-(d) have the additional input $m_{\frac{1}{2}}=140$.
(a) and (c) are for branching ratios at the tree level while 
(b) and (d) include the loop correction.
All masses are in unit of GeV and angles in unit of radian.}                       
\label{newtrilep} 
\end{figure}

An interesting phenomena arises if $H^{-}$ decays into a 
$\chi^{-}_1\chi_2^0$. The subsequent decays of $\chi^{-}_1$ and
$\chi_2^0$ can produce a trileptonic signal 
$H^{-}\rightarrow \chi^{-}_1 \chi_2^0$ 
$\rightarrow$ ${\it l_1^{-}{\it l_2^+} {\it l}_2^-}$.
Such a signal is well known in the context of the decay of 
the W boson. For on shell decays it was discussed
in early works\cite{onshell} and in off shell decays in Ref.\cite{Nath:sw}.
(For a recent analysis see Ref.\cite{sugratrilep}).
For the charged Higgs here, the signal can appear for on shell decays
since the mass of the Higgs is expected to
be large enough for such a decay to occur on shell.
In Fig.~(\ref{newtrilep}) we give an analysis of the branching ratio
of $H^{-}$ decay into $\chi^{-}_1\chi_2^0$ which enters in the
trileptonic signal. Plots as a function of $m_{\frac{1}{2}}$ 
 (Fig.~\ref{fig17a} - Fig.~\ref{fig17b}) 
 and as a function of $m_0$ (Fig.~\ref{fig17c} - Fig.~\ref{fig17d}) 
  are given where Fig.~\ref{fig17a} - Fig.~\ref{fig17c}
  are at the tree level and Fig.~\ref{fig17b} - Fig.~\ref{fig17d}
include the loop correction. A comparison of 
Fig.~\ref{fig17a} and  Fig.~\ref{fig17b}
and of   Fig.~\ref{fig17c} and Fig.~\ref{fig17d}
 shows that the loop correction to the
branching ratio is quite substantial up to  20-30\%.
Thus one may expect  the supersymmetric radiative
correction to the trileptonic signal to be substantial 
reaching up to the level of $20-30\%$.
However, a  full analysis of the
loop correction on the trileptonic signal would require an
analysis of the loop correction to the various decay modes of
the charginos and neutlinos. Such an analysis is outside the
scope of this paper and requires an independent study. 

\section{Conclusion}
In this paper we have carried out an analysis of the supersymmetric loop 
correction to the  $\chi^{-}\chi^0H^+$ couplings within MSSM.
This analysis extends previous analyses where supersymmetric loop correction 
to the couplings $\bar t bH^+$ and 
$\tau^- \bar\nu_{\tau}H^+$  within the minimal 
supersymmetric standard model including the full set of allowed 
CP phases. The result of the analysis is then applied to the computation of 
the decay of the charged Higgs $H^-$ to  $\bar t b$,  $\tau \bar \nu_{\tau}$,  
and $\chi_i^{-}\chi_j^0$ (i=1,2; j=1-4). The effect of the supersymmetic
loop ocrrection is found to be rather large, as much as 20-30\% in
significant regions of the parameter space. Further, 
the supersymmetric loop correction is found to be sizable for 
the full set of decay modes. Specific attention is paid to the 
chargino-neutralino decay mode that can lead to a trileptonic signal.
It is found that the effect on these modes can also be significant
reaching as much 20-30\% and thus the trileptonic signal would be
affected at this level. The effect of CP phases on the loop correction
are also investigated and it is found that the loop correction was
indeed very sensitive to the phases and that CP effects can affect
the loop correction significantly consistent with the edm constraints.

\noindent
{\bf Acknowledgments}\\ 
This research was also supported in part by NSF grant PHY-0139967.\\

\newpage


\newpage

\newpage



\end{document}